\documentclass[prb,twocolumn,groupedaddress,shownopacs,amssymb]{revtex4-2}
\usepackage{graphicx,psfrag,amsmath,amssymb}

\usepackage[usenames, dvipsnames]{color}

\usepackage{hyperref} 
\hypersetup{
    colorlinks=true,
    linkcolor=Blue,
    citecolor=Blue,
    filecolor=Blue,      
    urlcolor=Blue,
    pdftitle={pyrochlore},
    }

\begin{document}

\title{Cooper pairing, flat-band superconductivity and quantum geometry 
\\ in the pyrochlore-Hubbard model}

\author{M. Iskin}
\affiliation{
Department of Physics, Ko\c{c} University, Rumelifeneri Yolu, 
34450 Sar\i yer, Istanbul, T\"urkiye
}

\date{\today}

\begin{abstract}

We investigate the impacts of the quantum geometry of Bloch states, specifically 
through the band-resolved quantum-metric tensor, on Cooper pairing and 
flat-band superconductivity in a three-dimensional pyrochlore-Hubbard model. 
First we analyze the low-lying two-body spectrum exactly, and show that the 
pairing order parameter is uniform in this four-band lattice. This allowed us 
to establish direct relations between the superfluid weight of a multiband 
superconductor and ($i$) the effective mass of the lowest-lying two-body 
branch at zero temperature, ($ii$) the kinetic coefficient of the 
Ginzburg-Landau theory in proximity to the critical temperature, and 
($iii$) the velocity of the low-energy Goldstone modes at zero temperature. 
Furthermore, we perform a comprehensive numerical analysis of the superfluid 
weight and Goldstone modes, exploring both their conventional and geometric 
components at zero temperature. 

\end{abstract}

\maketitle

\section{Introduction}
\label{sec:intro}

The complex quantum geometric tensor serves as a central and defining 
concept in modern solid-state and condensed-matter 
physics~\cite{Provost80, berry84, resta11}. Its imaginary 
component is in the form of an anti-symmetric tensor known as the 
Berry curvature, and the associated Chern number has proven instrumental 
in the classification of topological insulators and 
superconductors~\cite{xiao10, qi11, chiu16, bansil16}. 
Its real part is in the form of a symmetric tensor known as the quantum 
metric, and it quantifies the quantum distance between adjacent Bloch states. 
Despite a long history of interdisciplinary interest in various physical 
phenomena governed by the Berry curvature, nature has been less forthcoming 
regarding the effects of the quantum metric. Only in the past decade or 
so have researchers increasingly recognized the significance of the 
quantum metric in various contexts. Notably, following the pioneering work 
by Peotta and T\"orm\"a in 2015 on the origins of superfluidity in 
topologically-nontrivial flat bands~\cite{peotta15}, a deeper connection 
between the transport properties of a multiband superconductor and the 
quantum geometry of its Bloch states has emerged as a surprising revelation 
in recent years~\cite{torma22, huhtinen22, torma23, hu19, herzog22a, hofmann23}.

Theoretical interest in flat-band superconductivity dates back a long time, as 
materials hosting quasi-flat Bloch bands were envisioned as a potential pathway 
to achieve room-temperature superconductivity~\cite{khodel90, kopnin11}. 
This anticipation was grounded in the naive BCS theory, which was suggested 
by the relatively elevated single-particle density of states within narrower 
Bloch bands. However, it is crucial to emphasize that the microscopic mechanism 
underpinning the emergence of flat-band superconductivity was completely absent 
in these earlier studies. 
It remained unclear whether superconductivity could thrive within an isolated 
flat band, given that the infinite effective band mass hampers the potential 
for localized particles on the lattice to attain superconductivity, thus acting 
as an inhibiting factor. As a result, it was believed that superconductivity was 
strictly prohibited when the permissible Bloch states originated solely from a 
single flat band~\cite{classen20}.

Recent studies illuminated these two perplexing arguments and unveiled a 
physical mechanism that theoretically permits the existence of 
flat-band superconductors~\cite{torma22, huhtinen22, torma23}. 
It has been demonstrated that multiband lattices (such as in Moir\'e materials) 
contribute differently to the superfluid weight. The real intraband processes 
were associated with the conventional contribution, while the virtual interband 
processes were linked to the geometric aspect. 
Unlike the conventional contribution~\cite{scalapino92, denteneer93}, 
which is solely determined by the derivatives of the Bloch bands, 
the geometric contribution is also influenced by the derivatives of 
the associated Bloch states. Consequently, unless the 
geometric contribution is nullified, superconductivity can manifest within 
a flat band, thanks to the involvement of other flat or dispersive bands 
through interband processes. There are also alternative proposals involving 
two-band systems as a potential means to achieve high-critical temperatures 
in the BCS-BEC crossover.~\cite{tajima19, tajima20}.

These findings highlight the necessity of considering not only the dispersion of 
the Bloch bands but also the geometry of the Bloch states in the pursuit of 
high-critical-temperature superconductivity. It is only by incorporating both 
factors that we can fully exploit their potential. A non-trivial quantum geometry 
is indispensable, as the mere presence of a flat band does not guarantee 
superconductivity if its geometry is trivial.
Thus, the emerging field of quantum geometry within multiband superconductors 
holds substantial potential for advancing our understanding of flat-band 
superconductors, assuming Hubbard-type tight-binding Hamiltonians mimic their 
underlying low-energy physics. Despite the significant progress made with 
one-dimensional and two-dimensional lattices that feature flat bands
~\cite{julku16, liang17, iskin18c, iskin19, wu21, torma22, huhtinen22, 
chan22, chan22b, kitamura22, porlles23, chen23, jiang23, hu23}, 
there has been limited exploration of more realistic three-dimensional lattices 
due to the technical complexities and challenges associated with their analysis. 
Our aim is to fill this gap by investigating quantum-geometric effects in a
pyrochlore-Hubbard model, where the pyrochlore lattice consists of a 
three-dimensional arrangement of tetrahedra sharing corners, possesses cubic symmetry, 
and is commonly encountered in transition-metal and rare-earth oxide materials,
especially in oxide compounds~\cite{gardner10}.  Given the recent demonstrations 
of three-dimensional flat bands and superconductivity in a pyrochlore metal 
CaNi$_2$~\cite{wakefiel23} and pyrochlore superconductor CeRu$_2$~\cite{huang23},
these structures present a compelling lattice platform for exploring the interplay 
between quantum geometry and strong correlations.

The rest of the paper is organized as follows. In Sec.~\ref{sec:onebody} we introduce
the pyrochlore lattice, and discuss its one-body spectrum. In Sec.~\ref{sec:twobody}
we calculate the low-lying two-body spectrum for the pyrochlore-Hubbard model, 
and derive the effective mass tensor of the lowest-lying branch. 
In Sec.~\ref{sec:manybody} we analyze the superfluid weight at zero and finite 
temperatures, relate it to the velocity of the low-energy Goldstone modes at 
zero temperature, and present their thorough numerical exploration in 
Sec.~\ref{sec:numerics}. The paper ends with a summary and an outlook in
Sec.~\ref{sec:conc}, and the Gaussian fluctuations and numerical implementations 
are discussed in Appendices~\ref{sec:gf} and~\ref{sec:ni}.

\section{One-body problem}
\label{sec:onebody}

The pyrochlore lattice is one of the simplest three-dimensional tight-binding models 
that feature a flat band in the Bloch spectrum~\cite{guo09, mizoguchi19}. 
It has an underlying face-centered-cubic 
Bravais lattice that can be defined by the primitive unit vectors
$\mathbf{a}_1 = (0, a/2, a/2)$, $\mathbf{a}_2 = (a/2, 0, a/2)$ and
$\mathbf{a}_3 = (a/2, a/2, 0)$, where $a$ is the side-length of the conventional 
simple-cubic cell. Its basis consists of $N_S = 4$ sublattice sites that are 
located at $\mathbf{r}_A = (0, 0, 0)$, $\mathbf{r}_B = \mathbf{a}_1/2$, 
$\mathbf{r}_C = \mathbf{a}_2/2$ and $\mathbf{r}_D = \mathbf{a}_3/2$. The corresponding 
first Brillouin Zone (BZ) has the shape of a truncated octahedron with a side-length 
$\sqrt{2}\pi/a$. The associated reciprocal space is such that
$
\sum_{\mathbf{k} \in \mathrm{BZ}} 1 = N_c,
$
where $\mathbf{k} = (k_x, k_y, k_z)$ is the crystal momentum in units of $\hbar \to 1$
and $N_c$ is the number of unit cells in the system. That is the total volume of the 
system is $\mathcal{V} = N_c a^3/4$, where $32 \pi^3/a^3$ is the volume of the BZ in 
reciprocal space, $a^3/4$ is the volume of the primitive cell in real space, and
$N = 4 N_c$ is the total number of lattice sites in the system.

\begin{figure} [htb]
\centerline{\scalebox{0.45}{\includegraphics{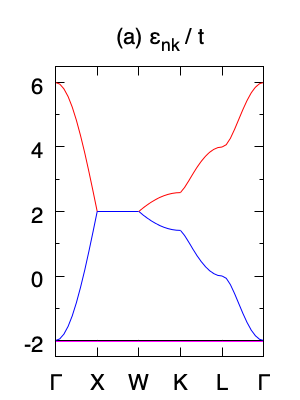} \includegraphics{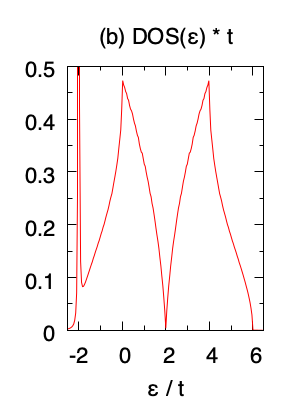}}}
\caption{\label{fig:onebody} 
(a) Bloch bands are shown along the high-symmetry points in the first BZ where
$\Gamma \equiv (0,0,0)$, $X \equiv (0,0,2\pi/a)$, $W \equiv (\pi/a,0,2\pi/a)$,
$K \equiv (\pi/2a,\pi/2a,2\pi/a)$, and $L \equiv (\pi/a,\pi/a,\pi/a)$.
(b) Density of states per unit cell per spin as a function of energy.
}
\end{figure}

Having a spin-$1/2$ system in mind with $\sigma = \{\uparrow, \downarrow\}$ labeling
the spin projections, the Bloch Hamiltonian for such a lattice can be written as
$
\mathcal{H}_0 = \sum_{S S' \sigma \mathbf{k}} h_{SS'}^{\mathbf{k}} 
c_{S \mathbf{k} \sigma}^\dagger c_{S' \mathbf{k} \sigma},
$
where $c_{S \mathbf{k} \sigma}$ annihilates a spin-$\sigma$ particle on the 
sublattice $S \equiv \{A, B, C, D\}$ with momentum $\mathbf{k}$. The elements
$h_{SS'}^\mathbf{k} = h_{S'S}^\mathbf{k}$ of the Hamiltonian matrix 
$\mathbf{h}_\mathbf{k}$ are real such that
$
h_{SS}^\mathbf{k} = 0,
$
$
h_{AB}^\mathbf{k} = -2\bar{t} \cos\big(\frac{k_y+k_z}{4}a\big),
$
$
h_{AC}^\mathbf{k} = -2\bar{t} \cos\big(\frac{k_x+k_z}{4}a\big),
$
$
h_{AD}^\mathbf{k} = -2\bar{t} \cos\big(\frac{k_x+k_y}{4}a\big),
$
$
h_{BC}^\mathbf{k} = -2\bar{t} \cos\big(\frac{k_x-k_y}{4}a\big),
$
$
h_{BD}^\mathbf{k} = -2\bar{t} \cos\big(\frac{k_x-k_z}{4}a\big)
$
and
$
h_{CD}^\mathbf{k} = -2\bar{t} \cos\big(\frac{k_y-k_z}{4}a\big),
$
where $\bar{t}$ is the hopping parameter between the nearest-neighbor sites.
Thus
$
h_{SS'}^{\mathbf{k}} = (h_{SS'}^{-\mathbf{k}})^*
$
respects time-reversal symmetry. The resultant eigenvalue problem, i.e.,
\begin{align}
\label{eqn:hSS}
\sum_{S'} h_{SS'}^\mathbf{k} n_{S' \mathbf{k}} = \varepsilon_{n\mathbf{k}} n_{S \mathbf{k}},
\end{align}
leads to four Bloch bands in the one-body spectrum, where
$
\varepsilon_{1\mathbf{k}} = -2\bar{t}(1 + \sqrt{1 + \Lambda_\mathbf{k}})
$
and
$
\varepsilon_{2\mathbf{k}} = -2\bar{t}(1 - \sqrt{1 + \Lambda_\mathbf{k}})
$
are the dispersive bands with
$
\Lambda_\mathbf{k} = \cos(k_x a/2) \cos(k_y a/2) + 
\cos(k_y a/2) \cos(k_z a/2) + \cos(k_x a/2) \cos(k_z a/2),
$
and
$
\varepsilon_{3\mathbf{k}} = \varepsilon_{4\mathbf{k}} = 2\bar{t}
$
are the flat bands. These bands are sketched in Fig.~\ref{fig:onebody}(a)
along the high-symmetry points. In this paper, since we prefer the flat bands 
to appear at the bottom of the spectrum, we set $\bar{t} \to -t$ and choose 
$t > 0$ as the unit of energy. 

In Fig.~\ref{fig:onebody}(b), we also show the single-particle density of 
states per unit cell per spin
$
DOS(\varepsilon) = \frac{1}{N_c} \sum_{n \mathbf{k}} 
\delta(\varepsilon - \varepsilon_{n\mathbf{k}}),
$
where the Dirac-delta function is represented via a Lorentzian distribution
$
\delta(x) = \frac{1}{\pi} \lim_{\eta \to 0} \frac{\eta}{x^2 + \eta^2}
$
with $\eta = 0.001t$. This is the origin of the energy broadening around 
$\varepsilon = -2t$. The van Hove singularities are clearly visible at 
$\varepsilon = \{ 0, 4 t\}$, and the density of states vanishes linearly at 
$\varepsilon = 2t$ with logarithmic corrections in its vicinity. The total 
bandwidth is $8t$.

\section{Two-body problem}
\label{sec:twobody}

\begin{figure} [htb]
\centerline{\scalebox{0.5}{\includegraphics{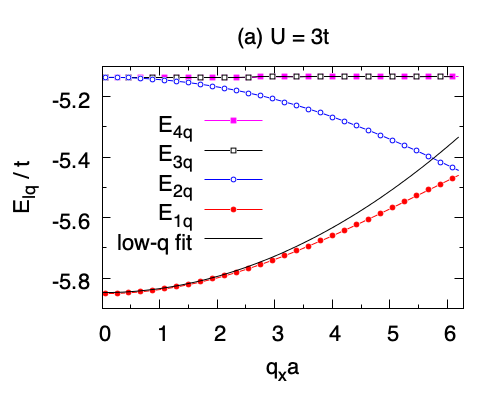}}}
\centerline{\scalebox{0.45}{\includegraphics{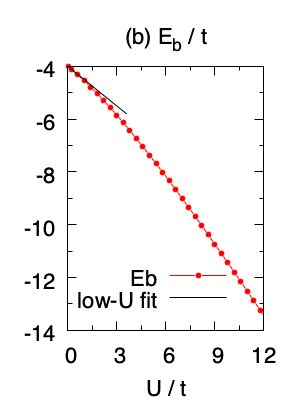} 
\hskip -0.8cm
\includegraphics{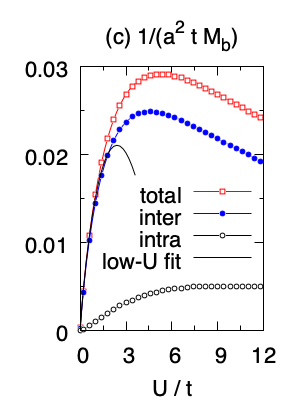}}}
\caption{\label{fig:twobody} 
(a) Spectrum $E_{\ell \mathbf{q}}$ of the low-lying bound states for $U = 3t$ 
as a function of $q_x$, when $q_y = 0$. The quadratic expansion
$E_{1 \mathbf{q}} = E_b + q_x^2/(2M_b)$ is an excellent fit for the lowest 
branch in the small-$q_x$ region.
(b) Energy offset $E_{1 \mathbf{0}}$ as a function of $U$. 
(c) Effective isotropic mass $1/M_b = 1/M_b^\mathrm{intra} + 1/M_b^\mathrm{inter}$ 
as a function of $U$ along with its intraband and interband contributions.
The low-$U$ fit
$
1/(a^2 t M_b) \approx  0.028 (U/t) - 0.013 (U/t)^{1.45}
$
is discussed in the text.
}
\end{figure}

In this paper we consider the simplest Hubbard model with an onsite attractive 
interaction between an $\uparrow$ and a $\downarrow$ particle,
\begin{align}
\label{eqn:Ham}
\mathcal{H} = \mathcal{H}_0 - \frac{U}{N_c} \sum_{S \mathbf{k} \mathbf{k'} \mathbf{q}} 
c_{S \mathbf{k} \uparrow}^\dagger c_{S, \mathbf{-k+q}, \downarrow}^\dagger 
c_{S, \mathbf{-k'+q}, \downarrow} c_{S \mathbf{k'} \uparrow},
\end{align}
where $U \ge 0$. The two-body problem can be solved exactly through a variational 
approach~\cite{iskin22}, leading to a number of spin-singlet bound states for 
a given center-of-mass momentum $\mathbf{q}$. It turns out the low-lying two-body 
spectrum $E_{\ell \mathbf{q}}$ can be determined by
$
\mathbf{G_q} \boldsymbol{\beta}_{\ell \mathbf{q}} = 0,
$
where $\mathbf{G_q}$ is a $4 \times 4$ Hermitian matrix with the following elements
\begin{align}
\label{eqn:GSS}
G_{SS'}^\mathbf{q} = \delta_{SS'} - \frac{U}{N_c} \sum_{n m \mathbf{k}}
\frac{m_{S, \mathbf{k-q}}^* n_{S \mathbf{k}} n_{S' \mathbf{k}}^* m_{S', \mathbf{k-q}}}
{\varepsilon_{n \mathbf{k}} + \varepsilon_{m, \mathbf{k-q}} - E_{\ell \mathbf{q}}}.
\end{align}
Here $\delta_{ij}$ is a Kronecker delta, and we assumed time-reversal symmetry 
$
\varepsilon_{n, -\mathbf{k} \downarrow} = \varepsilon_{n \mathbf{k} \uparrow}
= \varepsilon_{n \mathbf{k}}
$
and 
$
n_{S, -\mathbf{k} \downarrow}^* = n_{S \mathbf{k} \uparrow} = n_{S \mathbf{k}},
$ 
where $n_{S \mathbf{k}}$ is the sublattice projection $\langle S |n_\mathbf{k} \rangle$ 
of the periodic part $|n_\mathbf{k}\rangle$ of the Bloch state that is associated 
with $\varepsilon_{n \mathbf{k}}$ through Eq.~(\ref{eqn:hSS}). 
This yields a self-consistency relation for a given $E_{\ell \mathbf{q}}$, and its 
solutions can be found by setting the eigenvalues of $\mathbf{G_q}$ to $0$ one at 
a time. The corresponding eigenvectors
$
\boldsymbol{\beta}_{\ell \mathbf{q}} = (\beta_{A \ell \mathbf{q}}, 
\beta_{B \ell \mathbf{q}}, \beta_{C \ell \mathbf{q}}, \beta_{D \ell \mathbf{q}})^\mathrm{T}
$
can be used to characterize some physical properties of the bound states,
where $\mathrm{T}$ is the transpose.
Thus, for a given $\mathbf{q}$, there are 4 bound states below the threshold $-4t$ 
of the lowest two-body continuum, and we label them with $\ell = \{1, 2, 3, 4\}$ 
starting with the lowest branch. These solutions are illustrated in 
Fig.~\ref{fig:twobody}(a) for $U = 3t$, where the degenerate branches 
$E_{3\mathbf{q}} = E_{4\mathbf{q}}$ appear almost flat in the shown scale since 
their bandwidths are roughly $0.002t$. The overall structure of the two-body 
spectrum is reminiscent of the underlying Bloch spectrum but with the opposite 
sign of energy. Compare with the $\Gamma$ - $X$ portion in Fig.~\ref{fig:onebody}(a). 
This can be best understood in the $U/t \to \infty$, where the effective 
nearest-neighbor hopping parameter 
$t_p = 2t^2/U > 0$ for a strongly-bound pair of $\uparrow$ and $\downarrow$ 
particles has the opposite sign compared to $\bar{t} < 0$ of its unpaired 
constituents. This is because when a bound state breaks up at a cost of binding 
energy $U$ in the denominator and its $\uparrow$ constituent hops to a neighboring 
site, the $\downarrow$ partner follows it and hops to the same site, leading to 
$
\bar{t}_\uparrow \bar{t}_\downarrow = t^2
$ 
in the numerator. The prefactor accounts for the possibility of change in the order 
of spins. Such a virtual dissociation is the only physical mechanism for a 
strongly-bound pair of particles to move in the Hubbard model with 
nearest-neighbor hoppings.

Our numerical calculations also reveal that the so-called uniform-pairing 
condition~\cite{iskin22}, i.e.,
$
\boldsymbol{\beta}_{1 \mathbf{q}} \propto (1,1,1,1)^\mathrm{T},
$
is satisfied at all $U \ne 0$ for the lowest-lying $\ell = 1$ branch in the 
$\mathbf{q \to 0}$ limit. This finding suggests that the sublattice sites 
of a pyrochlore lattice in a unit cell must be equivalent by symmetry and make 
equal contribution to pairing. Thus, similar to the well-known two-dimensional 
toy models such as Mielke-checkerboard and kagome lattices 
that exhibit uniform pairing, the pyrochlore lattice offers an ideal playground 
for theoretical studies on flat-band superconductivity in three dimensions. 
For instance, when the uniform-pairing condition is met together 
with the underlying time-reversal symmetry, the energy $E_{1 \mathbf{q}}$ of the 
corresponding small-$\mathbf{q}$ bound states can be extracted simply from 
$\sum_{SS'} G_{SS'}^\mathbf{q} = 0$. In particular, it is possible to show 
that
\begin{align}
\label{eqn:E1q}
E_{1 \mathbf{q}} = E_b + \frac{1}{2}\sum_{ij} (M_b^{-1})_{ij} q_i q_j,
\end{align}
where the $\mathbf{q = 0}$ energy offset $E_b$ is determined by the self-consistency 
relation~\cite{iskin22}
\begin{align}
\label{eqn:Eb}
1 = \frac{U}{N} \sum_{n \mathbf{k}} \frac{1}{2\varepsilon_{n \mathbf{k}} - E_b}.
\end{align}
Furthermore, we split the elements of the inverse effective-mass tensor as
$
(M_b^{-1})_{ij} = (M_\mathrm{intra}^{-1})_{ij} + (M_\mathrm{inter}^{-1})_{ij},
$
depending on whether the intraband or interband processes are involved, 
leading to~\cite{iskin22}
\begin{align}
\label{eqn:Mintra}
(M_\mathrm{intra}^{-1})_{ij} &= \frac{
\sum_{n \mathbf{k}} \frac{\ddot{\varepsilon}_{n\mathbf{k}}^{ij}}{(2\varepsilon_{n \mathbf{k}} - E_b)^2}}
{2 \sum_{n \mathbf{k}} \frac{1}{(2\varepsilon_{n \mathbf{k}} - E_b)^2}}, \\
(M_\mathrm{inter}^{-1})_{ij} &= \frac{
\sum_{n \mathbf{k}} \frac{g_{ij}^{n \mathbf{k}}}{2\varepsilon_{n \mathbf{k}} - E_b}
- \sum_{n, m \ne n, \mathbf{k}} \frac{ g_{ij}^{nm\mathbf{k}}  }
{\varepsilon_{n \mathbf{k}} + \varepsilon_{m \mathbf{k}} - E_b} }
{\sum_{n \mathbf{k}} \frac{1}{(2\varepsilon_{n \mathbf{k}} - E_b)^2}}.
\label{eqn:Minter}
\end{align}
Here the intraband contribution depends only on the derivatives of the Bloch spectrum
$
\ddot{\varepsilon}_{n\mathbf{k}}^{ij} = \frac{\partial^2 \varepsilon_{n\mathbf{k}}}
{\partial k_i \partial k_j},
$
but the interband contribution depends also on the derivatives of the associated
Bloch states through the elements of the so-called band-resolved quantum-metric tensor
\begin{align}
\label{eqn:metric}
g_{ij}^{nm\mathbf{k}} = 2\mathrm{Re} \langle \dot{n}_\mathbf{k}^i | m_\mathbf{k} \rangle
\langle m_\mathbf{k} | \dot{n}_\mathbf{k}^j \rangle,
\end{align}
where $\mathrm{Re}$ denotes the real part and 
$
| \dot{n}_\mathbf{k}^i \rangle = \partial | n_\mathbf{k} \rangle/\partial k_i.
$ 
As the naming suggests, the elements of the quantum-metric tensor of the $n$th Bloch 
band~\cite{resta11} can be written as
$
g_{ij}^{n\mathbf{k}} = \sum_{m \ne n} g_{ij}^{nm\mathbf{k}}. 
$
Their origin can be traced back to the power-series expansion of
\begin{align}
\label{eqn:exp}
|\langle n_\mathbf{k} | m_\mathbf{k-q} \rangle|^2 
= \delta_{nm} - \sum_{ij} 
\frac{g_{ij}^{n\mathbf{k}} \delta_{nm} 
+ g_{ij}^{nm\mathbf{k}}(\delta_{nm} - 1)}{2} q_i q_j
\end{align}
in the small $\mathbf{q}$ limit.

In Figs.~\ref{fig:twobody}(b) and~\ref{fig:twobody}(c), we present the self-consistent 
solutions of Eqs.~(\ref{eqn:Eb}), (\ref{eqn:Mintra}) and~(\ref{eqn:Minter}) for the 
pyrochlore lattice, where the effective mass is isotropic in space. As an 
illustration, we show in Fig.~\ref{fig:twobody}(a) that the resultant 
$E_b \approx -5.8481t$ and $M_b \approx 37.3026 / (a^2 t)$ provide a perfect fit 
when $U = 3t$. Furthermore, in the $U/t \to 0$ limit when $E_b = -4t - U/2$, 
one can make analytical arguments that are based on some controlled approximations that 
$
1/(a^2 t M_b) \approx  \tilde{A}_1 (U/t) - \tilde{A}_2 (U/t)^{\tilde{A}_3} 
\gg 1/(a^2 t M_b^\mathrm{intra})
$
to the leading order, where $\tilde{A}_1$ and $\tilde{A}_2$ are numerical 
factors and $\tilde{A}_3 = 3/2$. 
Our numerical fit for $U/t \le 0.1$ shows that $\tilde{A}_1 \approx 0.028$, 
$\tilde{A}_2 \approx 0.013$ and $\tilde{A}_3 \approx 1.45$. 
This fit is shown in Fig.~\ref{fig:twobody}(c), and it works
quite well up to $U \lesssim t$.
Similarly, in the $U/t \to \infty$ limit when $E_b = - U$, 
one can show that $M_b = 4/(a^2 t_p)$ where $t_p = 2t^2/U$ is the
effective hopping parameter for a strongly-bound pair as discussed above.
This is consistent with our numerical finding $1/(a^2 t M_b) \approx 0.0048$ 
when $U = 100t$. It is pleasing to note that this finding is also aligned with 
the effective-mass tensor of the highest Bloch band where
$
(m_1^{-1})_{ij} = a^2 \ddot{\varepsilon}_{1 \mathbf{k}}^{ij}|_{\mathbf{k} = \mathbf{0}}
= (a^2 \bar{t}/4) \delta_{ij}.
$

\section{Many-body Problem}
\label{sec:manybody}

It turns out the complex parameter $\beta_{S \ell \mathbf{q}}$ plays precisely 
the role of an order parameter for spin-singlet pairing on sublattice $S$ in 
the two-body problem~\cite{iskin22}. In more general terms, the correspondence 
between the analogous parameters for the spin singlet and triplet two-body 
bound states and that of the spin singlet and triplet BCS order parameters can
be found in Ref.~\cite{iskin24b} in the context of the extended-Hubbard model. 
Thus, given that the uniform-pairing condition is satisfied for the 
lowest-lying two-body bound states when $\mathbf{q \to 0}$, 
the analogous Cooper pairing and its many-body BCS mean-field extension 
(i.e., assuming stationary Cooper pairs with zero center-of-mass momentum) 
are also described by a uniform order parameter in a unit cell. 
We emphasize that the uniform-pairing condition is satisfied exactly 
in the two-body problem without any phase difference between the sublattices.
This is in fact the underlying reason for the perfect agreement between 
Eqs.~(\ref{eqn:E1q})-(\ref{eqn:Minter}) and that of the numerics 
shown in Fig.~\ref{fig:twobody}(a).

\subsection{BCS-BEC crossover}
\label{sec:bcsbec}

For this reason, we take $\Delta_S \equiv \Delta_0$ as the uniform order parameter 
for superconductivity in all four sublattices, and set it to a real positive number. 
Following the standard prescription, we obtain the mean-field self-consistency 
relations~\cite{iskin18c}
\begin{align}
\label{eqn:op}
1 &= \frac{U}{N} \sum_{n\mathbf{k}} \frac{\mathcal{X}_{n\mathbf{k}}}{2E_{n\mathbf{k}}}, \\
F &= 1 - \frac{1}{N} \sum_{n\mathbf{k}} 
\frac{\mathcal{X}_{n\mathbf{k}}}{E_{n\mathbf{k}}} \xi_{n\mathbf{k}},
\label{eqn:filling}
\end{align}
where
$
\mathcal{X}_{n\mathbf{k}} = \tanh \big( \frac{E_{n\mathbf{k}}}{2 T} \big)
$
is a thermal factor with $k_B \to 1$ the Boltzmann constant and $T$ the temperature,
$
\xi_{n\mathbf{k}} = \varepsilon_{n\mathbf{k}} - \mu
$
is the shifted dispersion with $\mu$ the chemical potential,
$
E_{n\mathbf{k}} = \sqrt{\xi_{n\mathbf{k}}^2 + \Delta_0^2}
$
is the intraband quasiparticle spectrum, and the particle filling 
$0 \le F = \mathcal{N}/N \le 2$ corresponds to the total number of particles per site. 
Note that there is no interband pairing since the underlying time-reversal 
symmetry guarantees the presence of a $\downarrow$ particle in Bloch 
state $|n_{-\mathbf{k}}\rangle$ for every $\uparrow$ particle in $|n_\mathbf{k}\rangle$,
as the energetically most favorable BCS scenario for the stationary pairs.
Self-consistent solutions of Eqs.~(\ref{eqn:op}) and~(\ref{eqn:filling}) 
for $\Delta_0$ and $\mu$ is the starting point of the BCS-BEC crossover theories, 
and they are known to produce qualitatively correct results at sufficiently low $T$ 
including the $U/t \to \infty$ limit. In addition the mean-field expression for 
the filling of condensed particles~\cite{iskin18c, leggett}
\begin{align}
\label{eqn:Fc}
F_c = \frac{\Delta_0^2}{N} \sum_{n\mathbf{k}} \frac{\mathcal{X}_{n\mathbf{k}}^2}
{2E_{n\mathbf{k}}^2}
\end{align}
plays an important role in our discussion below, and it also produces qualitatively 
correct results at all $U \ne 0$ as long as $T$ is sufficiently low.

\subsection{Superfluid weight}
\label{sec:sfweight}

When the uniform-pairing condition is met together with the time-reversal symmetry,
the superfluid phase-stifness tensor or often called the superfluid weight can be 
written as
$
\mathcal{D}_{ij} = \mathcal{D}_{ij}^\mathrm{intra} + \mathcal{D}_{ij}^\mathrm{inter},
$
depending on whether the intraband or interband processes are involved. 
Using a linear-response theory and Kubo formalism or by simply imposing a 
phase-twist in the order parameter~\cite{liang17, iskin18c, daido23}, 
it can be shown that
\begin{align}
\label{eqn:Dijintra}
\mathcal{D}_{ij}^\mathrm{intra} &= \frac{\Delta_0^2}{\mathcal{V}} 
\sum_{n\mathbf{k}} 
\left(
\frac{\mathcal{X}_{n\mathbf{k}}}{E_{n\mathbf{k}}^3}
- \frac{\mathcal{Y}_{n\mathbf{k}}}{2T E_{n\mathbf{k}}^2}
\right) 
\dot{\xi}_{n\mathbf{k}}^i \dot{\xi}_{n\mathbf{k}}^j, \\
\mathcal{D}_{ij}^\mathrm{inter} &= \frac{\Delta_0^2}{\mathcal{V}} 
\sum_{n, m\ne n, \mathbf{k}} 
\left(
\frac{\mathcal{X}_{n\mathbf{k}}}{E_{n\mathbf{k}}}
- \frac{\mathcal{X}_{n\mathbf{k}}-\mathcal{X}_{m\mathbf{k}}}{E_{n\mathbf{k}}-E_{m\mathbf{k}}}
\right) \nonumber \\
&\hspace{2cm} \times \frac{(\xi_{n\mathbf{k}}-\xi_{m\mathbf{k}})^2} 
{E_{m\mathbf{k}}(E_{n\mathbf{k}}+E_{m\mathbf{k}})} g_{ij}^{nm\mathbf{k}},
\label{eqn:Dijinter}
\end{align}
where
$
\mathcal{Y}_{n\mathbf{k}} = \textrm{sech}^2\big( \frac{E_{n\mathbf{k}}}{2 T} \big)
$
is another thermal factor. Thus, similar to the inverse effective-mass tensor of 
the lowest-lying two-body branch, the intraband contribution depends only 
on the derivatives of the Bloch spectrum but the interband contribution depends 
also on the band-resolved quantum-metric tensor. For this reason the former 
(latter) is also referred to as the conventional (geometric) 
contribution~\cite{peotta15, liang17}. This is partly because Eq.~(\ref{eqn:Dijintra}) 
can be written in a more familiar form, i.e., through some integration by parts 
and algebra, 
$
\mathcal{D}_{ij}^\mathrm{intra} = \frac{\Delta_0^2}{\mathcal{V}} 
\sum_{n\mathbf{k}} [
\ddot{\xi}_{n\mathbf{k}}^{ij}
(1 - \mathcal{X}_{n\mathbf{k}} \xi_{n\mathbf{k}} / E_{n\mathbf{k}})
- \mathcal{Y}_{n\mathbf{k}} \dot{\xi}_{n\mathbf{k}}^i \dot{\xi}_{n\mathbf{k}}^j /(2T) ],
$
which is simply a sum over the the well-known single-band expression~\cite{denteneer93}.
Next we show that the geometric origin of the superfluid weight, i.e., 
Eq.~(\ref{eqn:Dijinter}), can be traced all the way back to the effective mass 
of the superfluid carriers.

\subsubsection{$T \to 0$ limit}
\label{sec:T0}

Lets first analyze Eqs.~(\ref{eqn:Dijintra}) and~(\ref{eqn:Dijinter}) at $T = 0$. 
Assuming $\Delta_0 \ne 0$, we may set
$
\mathcal{X}_{n\mathbf{k}} = 1
$
and
$
\mathcal{Y}_{n\mathbf{k}} = 0,
$
leading to
\begin{align}
\label{eqn:DijT0}
\mathcal{D}_{ij} = \frac{2\Delta_0^2}{\mathcal{V}} \sum_{n m \mathbf{k}} 
\frac{ \textrm{Re} [ 
\langle n_\mathbf{k}| \dot{\mathbf{h}}_\mathbf{k}^i |m_\mathbf{k} \rangle
\langle m_\mathbf{k}| \dot{\mathbf{h}}_\mathbf{k}^j | n_\mathbf{k} \rangle ] }
{E_{m\mathbf{k}}(E_{n\mathbf{k}}+E_{m\mathbf{k}})}
\end{align}
for the total superfluid weight, where 
$
\dot{\mathbf{h}}_\mathbf{k}^i = \partial \mathbf{h}_\mathbf{k} / \partial k_i
$
is the derivative of the Hamiltonian matrix. Here we concentrate on two physically 
transparent limits. The first one is the $U/t \to \infty$ limit, where
$
\Delta_0 = \frac{U}{2}\sqrt{F(2-F)}
$
and
$
\mu = -\frac{U}{2} (1-F)
$
such that
$
\sqrt{\mu^2 + \Delta_0^2} = U/2.
$
For this reason, we may set
$
E_{n \mathbf{k}} \to \sqrt{\mu^2 + \Delta_0^2}
$
in Eq.~(\ref{eqn:DijT0}), and calculate
$
\sum_{nm \mathbf{k}} [ 
\langle n_\mathbf{k}| \dot{\mathbf{h}}_\mathbf{k}^i | m_\mathbf{k} \rangle
\langle m_\mathbf{k}| \dot{\mathbf{h}}_\mathbf{k}^j | n_\mathbf{k} \rangle ]
= \sum_\mathbf{k} \mathrm{Tr}[\dot{\mathbf{h}}_\mathbf{k}^i 
\dot{\mathbf{h}}_\mathbf{k}^j]
= N_c a^2 t^2 \delta_{ij},
$
leading eventually to
$
\mathcal{D}_{ij} = \mathcal{D}_0 \delta_{ij}
$
where 
$
\mathcal{D}_0 = 8 F(2-F) t^2/(a U).
$
This result can be understood as follows. 
In the case of a continuum model with a single parabolic dispersion 
$\varepsilon_\mathbf{k} = k^2/(2m_0)$ and an attractive $s$-wave contact interaction 
between particles, it can be shown that $\mathcal{D}_0 = \rho_0/m_0$, 
where $\rho_0$ is the superfluid density. 
When these particles form strongly-bound weakly-interacting 
pairs in the BEC limit, we may set $\rho_p = \rho_0/2$ as the superfluid 
density of pairs and $M_p = 2 m_0$ as their mass, leading to 
$\mathcal{D}_0 = 4 \rho_p/M_p$ in terms of the pair 
properties~\cite{iskin18c, iskin19}. Furthermore, given 
that all of the particles participate in the superfluid flow at $T = 0$ for any
interaction strength in a continuum model~\cite{Fetter}, i.e.,
$
\rho_0 = \mathcal{N}/\mathcal{V} = 16 F/a^3
$
where $F$ is the filling of superfluid particles in the corresponding lattice model, 
we expect
$
\rho_p = 16 F_p/a^3
$
where $F_p$ is the filling of superfluid pairs.
Moreover, substituting $F_c/2 = F (2-F)/4$ as $F_p$ in the $U/t \to \infty$ limit, 
we identify
$
M_p = 2 U / (a^2 t^2)
$
as the effective mass of the superfluid pairs in a pyrochlore lattice.
It is pleasing to see that this analysis is in full agreement with 
$M_b = 2 U / (a^2 t^2)$ of the effective band-mass of the lowest-lying two-body 
branch presented in Sec.~\ref{sec:twobody}.

The second limit is the extremely-low particle-filling $F \to 0$ case and its
extremely-high particle-filling $F \to 2$ counterpart (which is equivalently to the 
extremely-low hole-filling), where $\mu \le -2t$ and $\xi_{n\mathbf{k}} \ge 0$ 
in the former, and $\mu \ge 6t$ and $\xi_{n\mathbf{k}} \le 0$ in the latter with 
$|\xi_{n\mathbf{k}}| \gg \Delta_0$ for any $U \ne 0$. For this reason we may set
$
E_{n \mathbf{k}} \to |\xi_{n\mathbf{k}}|,
$
and obtain
\begin{align}
\label{eqn:DijintralowF}
\mathcal{D}_{ij}^\mathrm{intra} &= \frac{\Delta_0^2}{\mathcal{V}} 
\sum_{n\mathbf{k}} 
\frac{\dot{\xi}_{n\mathbf{k}}^i \dot{\xi}_{n\mathbf{k}}^j}
{|\xi_{n\mathbf{k}}|^3}
= \frac{\Delta_0^2}{2 \mathcal{V}} \sum_{n\mathbf{k}} 
\frac{|\ddot{\xi}_{n\mathbf{k}}^{ij}|}
{\xi_{n\mathbf{k}}^2},
\\
\label{eqn:DijinterlowF}
\mathcal{D}_{ij}^\mathrm{inter} &= \frac{2\Delta_0^2}{\mathcal{V}} 
\sum_{n \mathbf{k}} \frac{g_{ij}^{n \mathbf{k}}}{|\xi_{n \mathbf{k}}|}
- \frac{4\Delta_0^2}{\mathcal{V}} 
\sum_{n, m \ne n, \mathbf{k}} \frac{g_{ij}^{nm\mathbf{k}}}
{ |\xi_{n \mathbf{k}}| + |\xi_{m \mathbf{k}}|},
\end{align}
where we used integration by parts in Eq.~(\ref{eqn:DijintralowF}), and
$
g_{ij}^{nm\mathbf{k}} = g_{ji}^{mn\mathbf{k}} 
$
and
$
g_{ij}^{n\mathbf{k}} = g_{ji}^{n\mathbf{k}} 
$
in the evaluation of Eq.~(\ref{eqn:DijinterlowF}). In addition, using the 
density of superfluid pairs
$
\rho_p = N F_p/\mathcal{V} 
$
along with
$
F_p = F_c/2 = \frac{\Delta_0^2}{4N} \sum_{n\mathbf{k}} \frac{1}{\xi_{n\mathbf{k}}^2},
$
we eventually find
\begin{align}
\mathcal{D}_{ij} = 4 \rho_p (M_b^{-1})_{ij}
\end{align}
for the $F \to 0$ limit, where 
$
(M_b^{-1})_{ij} = (M_\mathrm{intra}^{-1})_{ij} + (M_\mathrm{inter}^{-1})_{ij}
$
is determined precisely the inverse effective mass tensor of the lowest-lying 
two-particle bound states. This can be seen by setting $\mu = E_b/2$ and 
$
\ddot{\xi}_{n\mathbf{k}}^{ij} = \ddot{\varepsilon}_{n\mathbf{k}}^{ij}
$
in Eqs.~(\ref{eqn:Mintra}) and~(\ref{eqn:Minter}).
Note that $F \to 2$ limit is similar but with the appearance of inverse effective mass 
tensor of the highest-lying two-hole branch.

\subsubsection{$T \to T_c$ limit}
\label{sec:Tc}

Lets also analyze Eqs.~(\ref{eqn:Dijintra}) and~(\ref{eqn:Dijinter}) in the vicinity 
of the critical superconducting transition temperature $T_c$. Since $\Delta_0/t \to 0^+$ 
in the $T \to T_c$ limit from below, we may set
$
E_{n\mathbf{k}} \to |\xi_{n\mathbf{k}}|,
$
and obtain
\begin{align}
\label{eqn:DijintraTc}
\mathcal{D}_{ij}^\mathrm{intra} &= \frac{\Delta_0^2}{\mathcal{V}} 
\sum_{n \mathbf{k}} \bigg[\left( 
\frac{X_{n\mathbf{k}}}{2\xi_{n\mathbf{k}}^2} 
- \frac{Y_{n\mathbf{k}}}{4T\xi_{n\mathbf{k}}} 
\right) \ddot{\xi}_{n\mathbf{k}}^{ij} 
+\frac{X_{n\mathbf{k}} Y_{n\mathbf{k}}}{4 T^2\xi_{n\mathbf{k}}}
\dot{\xi}_{n\mathbf{k}}^i \dot{\xi}_{n\mathbf{k}}^j
\bigg],
\\
\label{eqn:DijinterTc}
\mathcal{D}_{ij}^\mathrm{inter} &= \frac{2\Delta_0^2}{\mathcal{V}} 
\sum_{n \mathbf{k}}
\frac{X_{n\mathbf{k}}}{\xi_{n\mathbf{k}}} g^{n\mathbf{k}}_{ij}
-  \frac{2\Delta_0^2}{\mathcal{V}} \sum_{n, m \ne n, \mathbf{k}} \frac{X_{n\mathbf{k}} + X_{m\mathbf{k}}}
{\xi_{n\mathbf{k}}+\xi_{m\mathbf{k}}} g^{nm\mathbf{k}}_{ij},
\end{align}
where
$
X_{n\mathbf{k}} = \tanh\big( \frac{\xi_{n\mathbf{k}}}{2 T} \big)
$
and
$
Y_{n\mathbf{k}} = \textrm{sech}^2\big( \frac{\xi_{n\mathbf{k}}}{2 T} \big).
$
Here we used integration by parts
$
\sum_\mathbf{k} X_{n\mathbf{k}} \dot{\xi}_{n\mathbf{k}}^i \dot{\xi}_{n\mathbf{k}}^j 
/ \xi_{n\mathbf{k}}^3
= \sum_\mathbf{k} Y_{n\mathbf{k}} \dot{\xi}_{n\mathbf{k}}^i \dot{\xi}_{n\mathbf{k}}^j
/ (4 T \xi_{n\mathbf{k}}^2)
+ \sum_\mathbf{k} X_{n\mathbf{k}} \ddot{\xi}_{n\mathbf{k}}^{ij} / (2\xi_{n\mathbf{k}}^2)
$
that is followed by another integration by parts
$\sum_\mathbf{k} Y_{n\mathbf{k}} \dot{\xi}_{n\mathbf{k}}^i \dot{\xi}_{n\mathbf{k}}^j
/ \xi_{n\mathbf{k}}^2
= - \sum_\mathbf{k} X_{n\mathbf{k}} Y_{n\mathbf{k}} 
\dot{\xi}_{n\mathbf{k}}^i \dot{\xi}_{n\mathbf{k}}^j / (T \xi_{n\mathbf{k}})
+ \sum_\mathbf{k} Y_{n\mathbf{k}} \ddot{\xi}_{n\mathbf{k}}^{ij} / \xi_{n\mathbf{k}}
$
in Eq.~(\ref{eqn:DijintraTc}), and
$
g_{ij}^{nm\mathbf{k}} = g_{ji}^{mn\mathbf{k}} 
$
and
$
g_{ij}^{n\mathbf{k}} = g_{ji}^{n\mathbf{k}}
$
in the evaluation of Eq.~(\ref{eqn:DijinterTc}). 
Note that Eqs.~(\ref{eqn:DijintraTc}) and~(\ref{eqn:DijinterTc}) reproduce,
respectively, Eqs.~(\ref{eqn:DijintralowF}) and~(\ref{eqn:DijinterlowF}) in the 
limit when $|\xi_{n\mathbf{k}}| \gg T_c$, e.g., in the $U/t \to \infty$ where 
$T_c/t \propto t/U \to 0$.
It is also pleasing to see that
\begin{align}
\mathcal{D}_{ij} = \frac{4 N_c \Delta_0^2}{\mathcal{V}} \mathcal{C}_{ij},
\end{align}
where 
$
\mathcal{C}_{ij} = \mathcal{C}_{ij}^\mathrm{intra} + \mathcal{C}_{ij}^\mathrm{inter}
$ 
is precisely the kinetic coefficient that appears in the 
Ginzburg-Landau theory near $T_c$~\cite{iskin23}, determining not only 
the superfluid density and effective mass of the superfluid carriers but 
also the coherence length, magnetic penetration depth, upper critical magnetic field, etc.
Notably, certain among these quantities have already been measured to characterize 
geometric effects in twisted bilayer graphene~\cite{tian23}. 

We note in passing that it is desirable to have an analytic expression
for $\mathcal{C}_{ij}$ in the $U/t \to 0$ limit. However, due to the 
complex nature of the pyrochlore lattice, which features four Bloch 
bands with non-isolated flat bands touching dispersive bands, and 
dispersive bands exhibiting highly anisotropic momentum dependence 
within a nontrivial BZ that is in the shape of a truncated octahedron, 
evaluating $\mathcal{C}_{ij}$ poses a significant challenge. 
Moreover, numerically computing $\mathcal{C}_{ij}$ necessitates a 
self-consistent determination of $T_c$ and $\mu$, which, in principle, 
can be achieved by extending the Nozieres-Schmitt-Rink approach to 
the multiband case~\cite{nsr85, sademelo93}. Unfortunately, this 
extension is also highly nontrivial, and its numerical implementation 
may encounter challenging convergence issues. Therefore, we primarily 
focus on the $T = 0$ case in this paper, which is comparatively 
more manageable.

\subsection{Low-lying Goldstone modes at $T = 0$}
\label{sec:goldstone}

Similar to the superfluid weight, next we show that the low-energy 
collective modes also have a quantum-geometric origin~\cite{iskin20b}.
As discussed in the Appendix~\ref{sec:gf}, the dispersion $\omega_\mathbf{q}$ 
for the collective Goldstone modes is determined by the poles of the 
fluctuation propagator $\boldsymbol{\mathcal{M}}_q^{-1}$, i.e., by setting 
$\det \boldsymbol{\mathcal{M}}_q = 0$, after an analytic continuation 
$\mathrm{i}\nu_n \to \omega + \mathrm{i} 0^+$ to the real axis. At $T = 0$, the 
matrix elements of $\boldsymbol{\mathcal{M}}_q$ reduce to
\begin{align}
\mathcal{M}^{q,E}_{11} &= \frac{4}{U} + \frac{1}{N_c}
\sum_{nm \mathbf{k}} \frac{(\xi \xi' + E E')(E+E')
|\langle n_\mathbf{k} | m_{\mathbf{k}-\mathbf{q}} \rangle|^2}
{2 E E'[\omega^2 - (E+E')^2]}, \\
\mathcal{M}^{q,O}_{11} &= \frac{1}{N_c} 
\sum_{nm \mathbf{k}} \frac{(\xi E' + E \xi') \omega 
|\langle n_\mathbf{k} | m_{\mathbf{k}-\mathbf{q}} \rangle|^2}
{2 E E'[\omega^2 - (E+E')^2]}, \\
\mathcal{M}^q_{12} &= -\frac{1}{N_c} 
\sum_{nm \mathbf{k}} \frac{\Delta_0^2(E+E')
|\langle n_\mathbf{k} | m_{\mathbf{k}-\mathbf{q}} \rangle|^2}
{2 E E'[\omega^2 - (E+E')^2]},
\end{align}
where we denote $\xi_{n \mathbf{k}}$ by $\xi$, 
$\xi_{m, \mathbf{k}-\mathbf{q}}$ by $\xi'$,
$E_{n \mathbf{k}}$ by $E$ and
$E_{m, \mathbf{k}-\mathbf{q}}$ by $E'$. To determine the lowest-energy 
Goldstone modes, it is sufficient to retain terms up to quadratic 
order in their small $\mathbf{q}$ and $\omega$ expansions, leading to 
$
\mathcal{M}^{q,E}_{11} + \mathcal{M}^q_{12} = A + \sum_{ij} C_{ij} q_i q_j - D\omega^2,
$
$
\mathcal{M}^{q,E}_{11} - \mathcal{M}^q_{12} = \sum_{ij} Q_{ij} q_i q_j - R\omega^2
$
and
$
\mathcal{M}^{q,O}_{11} = -B\omega.
$
In the expansion of $\mathcal{M}^{q,E}_{11} - \mathcal{M}^q_{12}$, 
the zeroth-order term vanishes due to the saddle-point condition given
in Eq.~(\ref{eqn:op}). The non-kinetic expansion coefficients
$
A = \frac{1}{N_c} \sum_{n \mathbf{k}} \Delta_0^2/(2E_{n \mathbf{k}}^3),
$
$
B = \frac{1}{N_c} \sum_{n \mathbf{k}} \xi_{n \mathbf{k}}/(4E_{n \mathbf{k}}^3),
$
$
D = \frac{1}{N_c} \sum_{n \mathbf{k}} \xi_{n \mathbf{k}}^2/(8E_{n \mathbf{k}}^5)
$
and
$
R = \frac{1}{N_c} \sum_{n \mathbf{k}} 1/(8E_{n\mathbf{k}}^3)
$
are simply given by a sum of their single-band counterparts~\cite{engelbrecht97}.
When $B^2 \ll A R$, it is clearly seen that the phase and 
amplitude modes are decoupled, and this is known to be the case only 
in the strict BCS limit~\cite{engelbrecht97}.

Similar to the superfluid weight, the kinetic coefficients can be written as
$
Q_{ij} = Q_{ij}^\mathrm{intra} + Q_{ij}^\mathrm{inter}
$
and 
$
C_{ij} = C_{ij}^\mathrm{intra} + C_{ij}^\mathrm{inter},
$
depending on whether the intraband or interband processes are involved, 
leading to
\begin{align}
Q_{ij}^\mathrm{intra} &= \frac{1}{N_c} \sum_{n \mathbf{k}} 
\frac{1}{8 E_{n\mathbf{k}}^3}
\dot{\xi}_{n\mathbf{k}}^i \dot{\xi}_{n\mathbf{k}}^j, \\
Q_{ij}^\mathrm{inter} &= \frac{1}{N_c} \sum_{n, m\ne n, \mathbf{k}} 
\frac{(\xi_{n\mathbf{k}}-\xi_{m\mathbf{k}})^2}
{8 E_{n\mathbf{k}} E_{m\mathbf{k}} (E_{n\mathbf{k}}+E_{m\mathbf{k}})}
g^{nm\mathbf{k}}_{ij}, \\
C_{ij}^\mathrm{intra} &= \frac{1}{N_c} \sum_{n \mathbf{k}} 
\frac{1}{8 E_{n\mathbf{k}}^3} 
\left( 1 - \frac{5\Delta_0^2 \xi_{n\mathbf{k}}^2}{E_{n\mathbf{k}}^4} \right)
\dot{\xi}_{n\mathbf{k}}^i \dot{\xi}_{n\mathbf{k}}^j, \\
C_{ij}^\mathrm{inter} &= 
- \frac{1}{N_c} \sum_{n \mathbf{k}} 
\frac{\Delta_0^2} {4 E_{n\mathbf{k}}} g^{n\mathbf{k}}_{ij} \nonumber \\
&+ \frac{1}{N_c} \sum_{n, m\ne n, \mathbf{k}} 
\frac{(\xi_{n\mathbf{k}}-\xi_{m\mathbf{k}})^2 + 4\Delta_0^2}
{8 E_{n\mathbf{k}} E_{m\mathbf{k}} (E_{n\mathbf{k}}+E_{m\mathbf{k}})}
g^{nm\mathbf{k}}_{ij}.
\label{eqn:Cinter}
\end{align}
Here the geometric contributions follow from the small-$\mathbf{q}$ 
expansion given in Eq.~(\ref{eqn:exp})
~\footnote{
In the case of two-band lattices, our expansion coefficients recover all 
of the previous results except for the first line of Eq.~(\ref{eqn:Cinter}), 
which is missing there~\cite{iskin20b}.
}.
Furthermore, the intraband coefficients can also be put in the more 
familiar forms, i.e, through again some integration by parts and algebra,
$
Q_{ij}^\mathrm{intra} = \frac{1}{N_c} \sum_{n \mathbf{k}} 
[ \xi_{n \mathbf{k}} \ddot{\xi}_{n\mathbf{k}}^{ij}
- \dot{\xi}_{n\mathbf{k}}^i \dot{\xi}_{n\mathbf{k}}^j 
(1- 3 \Delta_0^2 / E_{n\mathbf{k}}^2) ]/(8 E_{n\mathbf{k}}^3)
$
and 
$
C_{ij}^\mathrm{intra} = \frac{1}{N_c} \sum_{n\mathbf{k}} 
[ \xi_{n\mathbf{k}} (1 - 3\Delta_0^2/E_{n\mathbf{k}}^2) \ddot{\xi}_{n\mathbf{k}}^{ij}
- \dot{\xi}_{n\mathbf{k}}^i \dot{\xi}_{n\mathbf{k}}^j 
(1- 10 \Delta_0^2 \xi_{n\mathbf{k}}^2 / E_{n\mathbf{k}}^4) ]/(8 E_{n\mathbf{k}}^3),
$
which are simply sums of their well-known single-band 
expressions~\cite{iskin20b, engelbrecht97}.
By setting $\det \boldsymbol{\mathcal{M}}_q = 0$, we obtain 
\begin{align}
\omega_\mathbf{q}^2 = \sum_{ij} \frac{Q_{ij}}{R+ B^2/A} q_i q_j,
\end{align}
which is the dispersion for the low-momentum and low-frequency Goldstone 
modes. Thus, we are pleased to verify that the low-energy 
collective excitations have a linear dispersion whose finite velocity is 
characterized by the superfluid weight, i.e.,
\begin{align}
\label{eqn:DQ}
\mathcal{D}_{ij} = \frac{8N_c \Delta_0^2}{\mathcal{V}} Q_{ij},
\end{align}
at zero temperature.

We note in passing that the stability of the Goldstone modes is ensured 
by the stability of the superconducting state, as follows. In the 
long-wavelength limit, the effective action associated with the phase 
fluctuations of the order parameter can be expressed as
$
\mathcal{S}_{\theta} = \frac{1}{8} \int_0^{1/T} d\tau \int d^3\mathbf{r} 
\sum_{ij} \mathcal{D}_{ij} \dot{\theta}^i \dot{\theta}^j,
$
where $\mathbf{r} = (x,y,z)$ denotes the position in real space and
$
\dot{\theta}^i = \partial \theta(\mathbf{r}, \tau) / \partial r_i
$
controls the spatial variations of the phase~\cite{loktev01, liang17}.
By definition, this implies that $\mathcal{D}_{ij}$ determines the 
response of the superconducting system to a phase twist of the order 
parameter, i.e., the response of the thermodynamic potential to 
an infinitesimal superfluid flow~\cite{taylor06}. Therefore, the 
stability of a spatially-uniform superfluid necessitates a positive definite 
$\mathcal{D}_{ij}$, as a negative eigenvalue indicates that the 
superconducting state is unstable towards a spontaneously generated phase 
gradient, i.e., towards a spatially-nonuniform superfluid. Moreoever, 
a positive definite $\mathcal{D}_{ij}$ ensures a positive speed for the 
Goldstone modes through Eq.~(\ref{eqn:DQ}) which is in accordance with 
the Landau's criterion for superfluidity.

\begin{figure} [htb]
\centerline{\scalebox{0.45}{\includegraphics{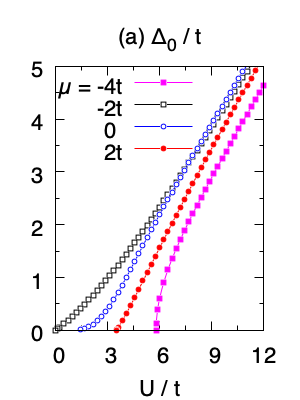} 
\hskip -0.8cm
\includegraphics{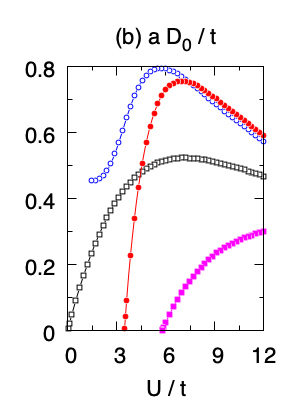}}}
\centerline{\scalebox{0.45}{\includegraphics{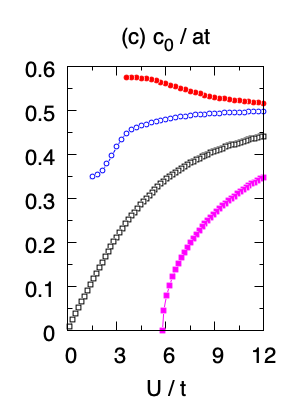}
\hskip -0.8cm
\includegraphics{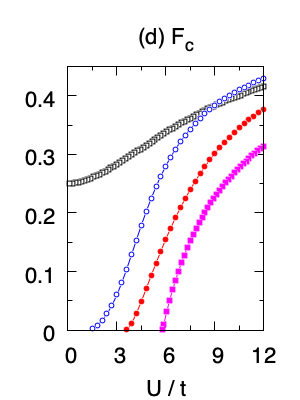}}}
\caption{\label{fig:gap} 
(a) Order parameter $\Delta_0$, (b) isotropic superfluid weight $\mathcal{D}_0$,
(c) isotropic sound speed $c_0$, and (d) filling of condensed particles $F_c$
as a function of $U$ at $T = 0$. 
Note that the disappearance of superconductivity is signalled by both 
$\Delta_0 \to 0$ and $\mathcal{D}_0 \to 0$ below the critical semi-metal 
point $U_c \approx 3.48t$ when $\mu = 2t$.
}
\end{figure}
\section{Numerical Results}
\label{sec:numerics}

In this section we exclusively set $T = 0$, and determine $\Delta_0$ and $\mu$ 
self-consistently from Eqs.~(\ref{eqn:op}) and~(\ref{eqn:filling}), and then 
plug them into Eqs.~(\ref{eqn:Dijintra}) and~(\ref{eqn:Dijinter}) as a 
function of $U$. Typical solutions are shown in Fig.~\ref{fig:gap}. 
When particle filling lies within the flat bands, i.e., when $\mu = -2t$ or 
equivalently $0 \le F \le 1$ at $U = 0$, it can be shown that~\cite{iskin17} 
$
\Delta_0 = \frac{U}{2} \sqrt{F(1-F)},
$
$
\mu = -2t - \frac{U}{2} \big( \frac{1}{2} - F \big)
$
and
$
F_c = F(1-F)
$
in the $U/t \to 0$ limit. Note that these expressions are quite similar to those
of the $U/t \to \infty$ limit's results, because, given that $U/W \gg 1$ with 
$W \to 0$ being the bandwidth of a flat band, even an arbitrarily small but 
finite $U \ne 0$ corresponds effectively to a strong-coupling limit. 
Accordingly, when $\mu = -2t$ coincides perfectly with a flat band at $U = 0$ 
(or equivalently corresponds to $F = 0.5$, i.e., to a half-filled flat bands), 
$\Delta_0$ grows linearly with $U$ as shown in Fig.~\ref{fig:gap}(a). 
In this case our numerical fit 
$
a \mathcal{D}_0/t =  \tilde{B}_1 (U/t) - \tilde{B}_2 (U/t)^{\tilde{B}_3}
$
for $U/t \le 0.1$ shows that $\tilde{B}_1 \approx 0.225$, 
$\tilde{B}_2 \approx 0.0544$ and $\tilde{B}_3 \approx 1.44$, 
and this fit works very well up to $U \lesssim 2t$.
This finding is in sharp contrast with the recent results on two-dimensional 
lattices, where a band touching with a flat band causes logarithmic corrections 
to the linear in $U$ term that is expected for an energetically-isolated flat 
band in any dimension~\cite{iskin19, wu21}.
It is pleasing to see that this fit is quite similar in structure to that 
of $1/M_b$ of the two-body problem discussed in Sec.~\ref{sec:twobody}, 
where $\tilde{B}_3 \approx \tilde{A}_3$. The ratios 
$\tilde{B}_1/\tilde{B}_2$ and $\tilde{A}_1/\tilde{A}_2$ are not expected to 
be similar unless $F \to 0$.

\begin{figure*}[htbp]
\includegraphics[scale=0.65]{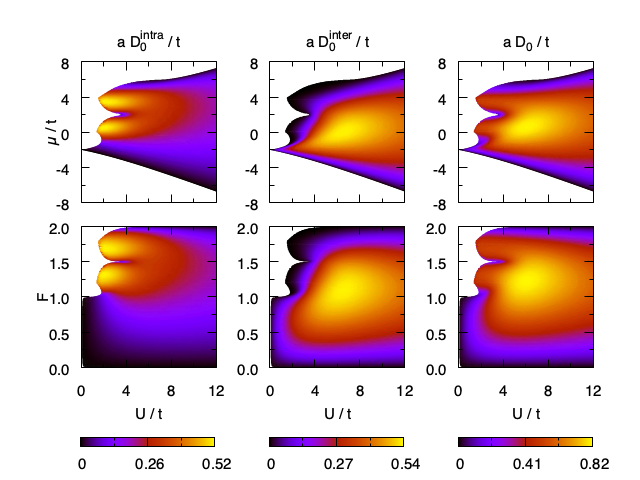}
\caption{
\label{fig:map}
Isotropic superfluid weight $\mathcal{D}_0$ together with its intraband and interband 
contributions as a function of $U$, $\mu$ and $F$ at $T = 0$. Since our 
numerical implementation becomes unreliable in the region where $\Delta_0$ 
is small, we choose to present those data that have $\Delta_0 \ge 0.01t$, 
which reveals the underlying $DOS(\mu)$ at the periphery of the white regions.
}
\end{figure*}

It can also be shown that, while the BCS order parameter $\Delta_0$ grows 
exponentially $e^{-1/[U DOS(\mu)]}$ slow when $-2t < \mu < 2t$ and $2t < \mu < 6t$ 
lies within any of the dispersive Bloch bands, it grows linearly $U-U_c$ fast 
from the critical semi-metal point with $U_c \approx 3.48t$ when $\mu = 2t$ or equivalently 
$F = 3/2$, and with a square root $\sqrt{U-U_c}$ from the particle and hole vacuums 
when $\mu < -2t$ or $\mu > 6t$. For instance $U_c \approx 5.79t$ when $\mu = -4t$.
These are illustrated in Fig.~\ref{fig:gap}(a). 
Thus, the semi-metal state persists even at finite $U$ and superconductivity 
does not appear until a critical interaction threshold $U_c$. The corresponding 
$\mathcal{D}_0$ are shown in Fig.~\ref{fig:gap}(b), where it saturates at 
sufficiently small $U/t$ when $\mu$ lies within any of the dispersive Bloch bands, 
and vanishes otherwise. The latter finding again signals the disappearance of 
superconductivity at a finite interval $U < U_c$.

In Fig.~\ref{fig:map} we present maps of the superfluid weight $\mathcal{D}_0$ 
together with its $\mathcal{D}_0^\mathrm{intra}$ and $\mathcal{D}_0^\mathrm{inter}$ 
contributions as a function of $U$, $\mu$ and $F$. Due to technical difficulties, i.e., 
the accuracy of the numerical integration becomes unreliable when the exponential 
growth of $\Delta_0$ is not accurately captured by the non-linear solver due 
to the convergence problems, we choose to present the data in the parameter
regime where $\Delta_0 \ge 0.01t$. See Appendix~\ref{sec:ni} for more details.
This is why Fig.~\ref{fig:map} has white regions, e.g., in the lower panels, 
even though $F$ corresponds to a partially-filled 
dispersive Bloch band in the $U/t \to 0$ limit where $\mathcal{D}_0$ is known 
to saturate. This drawback offers the advantage that the single-particle density 
of states $DOS(\mu)$ shown in Fig.~\ref{fig:onebody}(b) 
appears on the periphery of the white regions, including the flat bands 
at $\mu = -2t$, van Hove singularities at $\mu = \{0, 4t\}$ or equivalently 
at $F \approx \{1.19, 1.81\}$, and the critical semi-metal point at $\mu = 2t$ 
or equivalently at $F = 3/2$. On the other hand, the peripheries of the particle 
and hole vacuums and the critical semi-metal point are determined quite accurately
in the upper panels, since $\Delta_0$ vanishes very rapidly in their vicinity when 
$U \to U_c \ne 0$, e.g., see Fig.~\ref{fig:gap}(a). This is also indicated by the 
nearly invisible white regions at the edges of the lower panels when $F \to 0$ 
or $F \to 2$. 

\begin{figure}[htbp]
\includegraphics[scale=0.38]{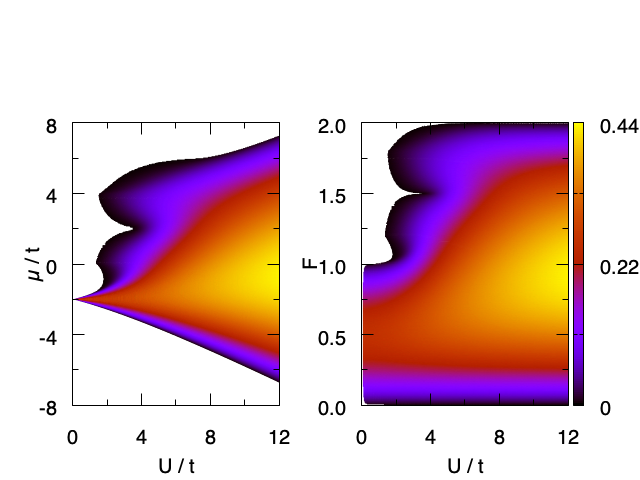}
\caption{
\label{fig:condpair}
The filling of condensed particles $F_c$ as a function of $U$, $\mu$ and $F$
at $T = 0$.
}\end{figure}

In the small $U/t$ regime, Fig.~\ref{fig:map} clearly shows that while 
$\mathcal{D}_0^\mathrm{inter}$ is the dominant contributor in the 
flat-band superconductivity, i.e., when $0 < F \le 1$, 
$\mathcal{D}_0^\mathrm{intra}$ dominates the usual superconductivity 
in general when $\mu$ lies within a dispersive band. However, both contributions 
are equally important in the large $U/t$ regime including the $U/t \to \infty$ 
limit (not shown). In fact 
$
\mathcal{D}_0^\mathrm{inter} \sim 2.6 \mathcal{D}_0^\mathrm{intra}
$ 
even at $U = 100t$. We checked that the total superfluid weight approaches 
$\mathcal{D}_0 = 8 F(2-F) t^2/(a U)$ in the $U/t \to \infty$ limit, which is 
in perfect agreement with the analysis given in Sec.~\ref{sec:T0}.
Since the filling of condensed particles $F_c$ plays a critical role
in connecting $\mathcal{D}_0$ to the effective mass of the superfluid carriers, we also 
present its map in Fig.~\ref{fig:condpair} as a function of $U$, $\mu$ and $F$.
This figure verifies that $F_c = F(1-F)$ (derived above) saturates when $0 \le F \le 1$ 
in the $U/t \to 0$ limit as soon as $U \ne 0$, and that all of the particles are 
entirely condensed in a dilute flat-band superconductor, i.e., $F_c = F$ when 
$F \to 0$ in Fig.~\ref{fig:condpair}. Such a perfect condensation may occur only 
if the repulsive interaction between small Cooper pairs is negligible, which may 
be the underlying reason behind our finding in Sec.~\ref{sec:T0} that
$
\mathcal{D}_{ij} = 4 \rho_p (M_b^{-1})_{ij}
$
is determined entirely by the effective mass of the lowest-lying two-body branch 
when $F \to 0$.
This is in sharp contrast with the usual superconductors where $F_c$ corresponds 
to a negligible fraction of particles in the small $U/t$ regime~\cite{leggett}, 
which can also be seen in the $F > 1$ region in Fig.~\ref{fig:condpair}. 
We again checked that $F_c = F(2-F)$ in the $U/t \to \infty$ limit, 
which is in perfect agreement with the analysis given in Sec.~\ref{sec:T0}.

\begin{figure*}[htbp]
\includegraphics[scale=0.65]{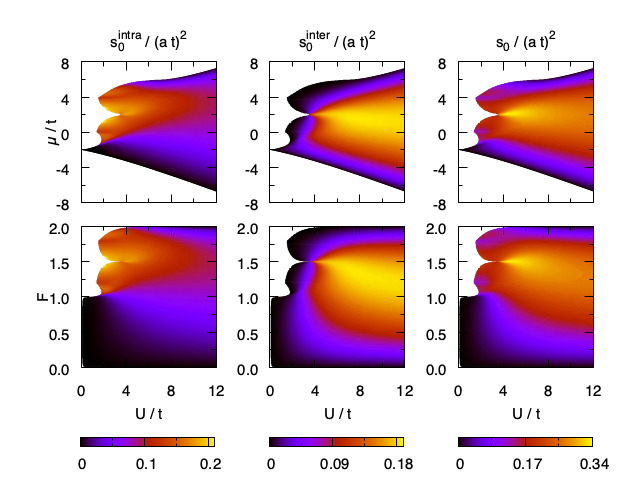}
\caption{
\label{fig:goldstone}
Square $s_0 = Q_0/(R+ B^2/A)$ of the isotropic sound speed together with 
its intraband and interband contributions as a function of $U$, $\mu$ and $F$ 
at $T = 0$. 
}
\end{figure*}

In Fig.~\ref{fig:goldstone} we present maps of the square 
$
s_0 = Q_0/(R + B^2/A)
$
of the isotropic sound speed $c_0$ together with its $s_0^\mathrm{intra}$ and 
$s_0^\mathrm{inter}$ contributions as a function of $U$, $\mu$ and $F$. 
First of all, both $s_0^\mathrm{intra}$ and $s_0$ saturate in the 
$U/t \to 0$ limit when $\mu$ lies within a dispersive band and 
$s_0^\mathrm{inter}$ is small and negligible. These are expected from 
the well-known single-band results~\cite{iskin07}. 
In addition, for a fixed $U/t$ in the small $U/t$ regime, both 
$s_0^\mathrm{intra}$ and $s_0$ exhibit faint but visible dips around 
the van Hove singularities, which are also consistent with the well-known 
single-band results~\cite{iskin07}. 
On the other hand, when $\mu = -2t$ coincides perfectly with a flat band at 
$U = 0$, both $s_0^\mathrm{inter}$ and $s_0$ grow from $0$ with a power-law 
(approximately quadratically) in $U$ but $s_0^\mathrm{intra}$ remain 
small and negligible. Thus, $c_0$ vanishes linearly with $U$ in a flat-band 
superconductor in the $U/t \to 0$ limit, which is in sharp contrast with 
the usual dispersive case where it saturates. 
In order to characterize and understand this particular limit,
we set $E_{n \mathbf{k}} \to \sqrt{(-2t-\mu)^2 + \Delta_0^2}$ for the 
flat $n = \{3,4\}$ bands and take only their contributions into account, 
leading to $A = 16F(1-F)/U$, $B = 8(1-2F)/U^2$ and $R = 16/U^3$, where
$0 \le F \le 1$. Furthermore, using the relation
$
\mathcal{D}_0 = 8 U^2 F(1-F) Q_0 /a^3,
$
we find $s_0 = U \mathcal{D}_0 a^3/32$, which is in perfect 
agreement with the numerics.
In addition, at the semi-metal critical point when $\mu = 2t$ and 
$U \to U_c$, both $s_0^\mathrm{intra}$
and $s_0^\mathrm{inter}$ exhibit comparable jumps from $0$. 
Because of this, the resultant $s_0$ exhibits a much larger jump right 
at the tip of the critical region, since $s_0^\mathrm{inter}$ is typically 
small and negligible in its vicinity.
Lastly we consider the $U/t \to \infty$ limit, and set 
$E_{n \mathbf{k}} \to \sqrt{\mu^2 + \Delta_0^2}$ for all bands,
leading to $A = 4F(2-F)/U$, $B = 4(1-F)/U^2$, $R = 4/U^3$ and 
$Q_0 = a^2 t^2/U^3$. The resultant $s_0 = a^2 t^2 F(2-F)/4$ 
depends only on $F$ as in the case of single-band case~\cite{iskin07}, 
and it is in perfect agreement with the numerics. 
It can be written in a form $s_0 = U \mathcal{D}_0 a^3/32$ that
is identical to the flat-band expression above.
We also find it instructive to reinterpret the sound speeed 
$c_0 = \sqrt{2U F_p/M_p}$ in terms of the filling $F_p$ and 
effective mass $M_p$ of the pairs. Then, by making a comparison with 
the Bogoliubov expression $c_B = \sqrt{U_B F_B / M_B}$ that is valid 
for a weakly-interacting Bose gas on a tight-binding lattice~\cite{rey03}, 
we identify $U_p = 2U$ as the parameter that characterizes the 
interaction between pairs in the $U/t \to \infty$ limit. Such a 
strong and repulsive onsite interaction between pairs of fermions 
can be attributed to the underlying Pauli exclusion principle~\cite{iskin07}. 
On a similar footing, using the results of Sec.~\ref{sec:T0} where 
$\mathcal{D}_0 = 4 \rho_p / M_b$ for a dilute ($F \to 0$) flat-band 
superconductor when $U/t \to 0$, we also reach the conclusion that 
$U_p = 2U$. This suggests that the interaction between pairs tends 
to zero in a flat-band superconductor.

\section{Conclusion}
\label{sec:conc}

To summarize, we studied the impact of the quantum geometry of Bloch states, 
specifically through the band-resolved quantum-metric tensor, on Cooper 
pairing and flat-band superconductivity within a three-dimensional 
pyrochlore-Hubbard model. For this purpose, first we showed that the pairing 
order parameter is uniform in this four-band lattice through an exact 
calculation of the low-lying two-body spectrum. This simplification enabled 
us to reveal direct relations between the superfluid weight of a multiband 
superconductor and ($i$) the effective mass of the lowest-lying two-body 
branch at $T = 0$ through 
$
\mathcal{D}_{ij} = 4 \rho_p (M_b^{-1})_{ij},
$
($ii$) the kinetic coefficient of the Ginzburg-Landau 
theory near $T_c$ through
$
\mathcal{D}_{ij} = \frac{4 N_c \Delta_0^2}{\mathcal{V}} \mathcal{C}_{ij},
$
and ($iii$) the velocity of the low-energy Goldstone modes at $T = 0$ through
$
\mathcal{D}_{ij} = \frac{8N_c \Delta_0^2}{\mathcal{V}} Q_{ij}.
$
The underlying physics behind these relations is that a bound state with 
a finite center-of-mass momentum is a collective mode of the superfluid ground 
state. Then we presented a thorough numerical analysis of the superfluid weight 
and Goldstone modes together with their intraband (conventional) and 
interband (geometric) contributions at zero temperature.
For instance, one of our important observations is that, in sharp contrast 
with the recent results on two-dimensional lattices where a band touching 
with a flat band causes logarithmic corrections 
$
\tilde{C}_1 U \ln(\tilde{C}_2/U)
$
to the superfluid weight in the $U/t \to 0$ limit, the analogous correction 
is a power law in three dimensions which may be approximated by 
$
\tilde{C}_1 U - \tilde{C}_2 U^{3/2}.
$
Another one is the relation $c_0 = \sqrt{U \mathcal{D}_0 a^3/32}$ 
between the sound speed and the superfluid weight in a flat-band 
superconductor when 
$
U/t \to 0,
$
which further suggests that the the interaction between pairs tends 
to zero.

Much like the well-explored two-dimensional toy models such as 
Mielke-checkerboard and kagome lattices, which display uniform pairing, we
believe the pyrochlore lattice presents an excellent setting for conducting 
theoretical research on three-dimensional flat-band superconductivity. 
As an outlook we are planning to develop a Ginzburg-Landau theory for the 
pyrochlore-Hubbard model, and explore how quantum geometry effects not only 
the superfluid density and effective mass of the superfluid carriers but 
also the coherence length, magnetic penetration depth, upper critical 
magnetic field, etc in the BCS-BEC crossover~\cite{iskin23}.

\begin{acknowledgments}
The author acknowledges funding from US Air Force Office of Scientific 
Research (AFOSR) Grant No. FA8655-24-1-7391.
\end{acknowledgments}

\appendix

\section{Gaussian Fluctuations}
\label{sec:gf}

In order to go beyond the saddle-point (mean-field) approximation, 
we use imaginary-time functional path-integral 
formalism~\cite{iskin20b, engelbrecht97} and derive 
a quadratic effective action in the fluctuations of the order parameter 
for a multiband Hubbard model with onsite attraction. It turns out this 
formalism is drastically simpler and transparent when the system 
manifests both time-reversal symmetry and uniform-pairing condition as 
in pyrochlore lattice of interest. In this case the bosonic 
Hubbard-Stratavonich field (which plays the role of the order parameter) 
can be split as
\begin{align}
\Delta_{S q} = \Delta_0 \delta_{q 0} + \Lambda_q
\end{align}
for all sublattice sites in a unit cell, where the complex field $\Lambda_q$ 
corresponds to the fluctuations around the stationary saddle-point parameter 
$\Delta_0$ (which is taken as real here and in the main text), and 
$
q \equiv (\mathbf{q}, \mathrm{i}\nu_\ell)
$ 
is a collective index with $\nu_\ell = 2\ell \pi T$ the bosonic Matsubara 
frequency. Given our interest in the low-energy in-phase (i.e., Goldstone) 
collective modes only, we also set $\Lambda_q$ to be uniform together with 
$\Delta_0$, making higher-energy out-of-phase (i.e., Leggett) collective 
modes inaccessible. In the case of a pyrochlore lattice, we expect three 
distinct Leggett branches to appear. In principle, they can be studied 
through an $S$dependent $\Lambda_{Sq}$, but this is beyond the scope of 
this paper. See~\cite{iskin20b} for a similar analysis in two-band 
lattices, showing that the speed of the Leggett modes also has an 
interband contribution that is controlled by the quantum metric. 

First we express the mean-field Hamiltonian $\mathcal{H}_\mathrm{mf}$ in 
the Bloch-band representation, whose Hamiltonian matrix can be expressed 
as~\cite{liang17}
\begin{align}
\mathbf{H}^\mathrm{mf}_\mathbf{k} = \sum_n 
(\xi_{n\mathbf{k}} \tau_z + \Delta_0 \tau_x)
\otimes |n_\mathbf{k} \rangle \langle n_\mathbf{k} |,
\end{align}
where $\tau_x$ and $\tau_z$ are Pauli matrices describing the particle-hole 
degrees of freedoms. Then the saddle-point propagator can be written as
$
\boldsymbol{\mathcal{G}}_0(k) =  \sum_{s n} \frac{| \Psi_{n \mathbf{k}}^s \rangle 
\langle \Psi_{n \mathbf{k}}^s |}
{\mathrm{i} \omega_\ell - E_{n \mathbf{k}}^s},
$
where 
$
k \equiv (\mathbf{k}, \mathrm{i}\omega_\ell)
$ 
is a combined index with $\omega_\ell = (2\ell+1) \pi T$ the fermionic Matsubara
frequency. Here
$
\mathbf{H}^\mathrm{mf}_\mathbf{k} | \Psi_{n \mathbf{k}}^s \rangle = 
E_{n \mathbf{k}}^s | \Psi_{n \mathbf{k}}^s \rangle
$
defines the quasiparticle-quasihole spectra 
$
E_{n \mathbf{k}}^s = s E_{n \mathbf{k}}
$
with $s = \pm$, where the associated eigenvectors are
$
| \Psi_{n \mathbf{k}}^+ \rangle = 
\left(
\begin{array}{c}
 u_{n \mathbf{k}} \\ v_{n \mathbf{k}} 
\end{array} 
\right) \otimes |n_\mathbf{k} \rangle
$
for the quasiparticles and
$
| \Psi_{n \mathbf{k}}^- \rangle = 
\left(
\begin{array}{c}
 -v_{n \mathbf{k}} \\ u_{n \mathbf{k}} 
\end{array} 
\right) \otimes |n_\mathbf{k} \rangle
$
for the quasiholes. The coherence factors
$
u_{n \mathbf{k}} = \sqrt{\frac{1}{2} + \frac{\xi_{n\mathbf{k}}}{E_{n\mathbf{k}}}}
$
and
$
v_{n \mathbf{k}} = \sqrt{\frac{1}{2} - \frac{\xi_{n\mathbf{k}}}{E_{n\mathbf{k}}}}
$
coincide with the usual intraband expressions due to the absence of 
interband pairing.

Following the standard procedure, the quadratic effective action can be written as
$
\mathcal{S}_2 = \frac{N_S}{T U} \sum_q |\Lambda_q|^2
+ \frac{1}{2 N_c} \mathrm{Tr} \sum_{kq} \boldsymbol{\mathcal{G}}_0(k) 
\mathbf{\Sigma}(q) \boldsymbol{\mathcal{G}}_0(k-q) \mathbf{\Sigma}(-q),
$
where $\mathrm{Tr}$ denotes a trace over the band and particle-hole 
sectors, and 
$
\boldsymbol{\Sigma}(q) = (\Lambda_q \tau_+ + \Lambda_{-q}^* \tau_-) \otimes 
\mathcal{I}_{N_S} 
$
is controlled purely by the fluctuation fields. 
Here $N_S$ is the number of sublattices in a unit cell (which is 4 for the 
pyrochlore lattice), $\tau_\pm = (\tau_x \pm i\tau_y)/2$ and 
$\mathcal{I}_{N_S}$ is an $N_S \times N_S$ unit matrix in the band space. 
Upon evaluation of the trace and sum over the fermionic frequencies, 
we eventually obtain 
\begin{align}
\label{eqn:}
\mathcal{S}_2 = \frac{1}{2T} \sum_q \left( \Lambda_q^* \, \Lambda_{-q} \right)
\left( \begin{array}{cc}
\mathcal{M}_q^{11} & \mathcal{M}_q^{12} \\
\mathcal{M}_q^{21} & \mathcal{M}_q^{22}
\end{array} \right)
\left( \begin{array}{c}
\Lambda_q \\ \Lambda_{-q}^*
\end{array} \right),
\end{align}
where the fluctuation matrix $\boldsymbol{\mathcal{M}}_q$ plays the role of 
inverse-propagator of amplitude and phase fluctuations. In order to 
express its matrix elements in a compact form, 
we denote $\xi_{n \mathbf{k}}$ by $\xi$,
$\xi_{m, \mathbf{k}-\mathbf{q}}$ by $\xi'$,
$E_{n \mathbf{k}}$ by $E$, $E_{m, \mathbf{k}-\mathbf{q}}$ by $E'$,
and denote their functions as
$u^2 = (1+\xi/E)/2$, $u'^2 = (1+\xi'/E')/2$, 
$v^2 = (1-\xi/E)/2$, $v'^2 = (1-\xi'/E')/2$, 
$f = 1/(e^{E/T}+1)$ and $f' = 1/(e^{E'/T}+1)$, leading to
\begin{align}
\mathcal{M}^q_{11} &= \mathcal{M}^{-q}_{22} = \frac{N_S}{U} + \frac{1}{N_c} 
\sum_{nm \mathbf{k}} 
|\langle n_\mathbf{k} | m_{\mathbf{k}-\mathbf{q}} \rangle|^2 \nonumber \\
\times &\bigg[(1-f-f') \left( \frac{u^2 u'^2}{\mathrm{i}\nu_\ell-E-E'} 
- \frac{v^2 v'^2}{\mathrm{i}\nu_\ell+E+E'} \right) \nonumber \\
+ &(f-f') \left( \frac{v^2 u'^2}{\mathrm{i}\nu_\ell+E-E'} 
- \frac{u^2 v'^2}{\mathrm{i}\nu_\ell-E+E'} \right) 
\bigg], \\
\mathcal{M}^q_{12} &= \mathcal{M}^q_{21} = \frac{1}{N_c} \sum_{nm \mathbf{k}} 
|\langle n_\mathbf{k} | m_{\mathbf{k}-\mathbf{q}} \rangle|^2 \nonumber \\
\times & \bigg[(1-f-f') \left( \frac{u v u 'v'}{\mathrm{i}\nu_\ell+E+E'} 
- \frac{u v u'v'}{\mathrm{i}\nu_\ell-E-E'} \right) \nonumber \\
+ &(f-f') \left( \frac{u v u' v'}{\mathrm{i}\nu_\ell+E-E'} 
- \frac{u v u' v'}{\mathrm{i}\nu_\ell-E+E'} \right) 
\bigg].
\end{align}
Note that while $\mathcal{M}^q_{12}$ is even both under 
$\mathbf{q} \to -\mathbf{q}$ and $\mathrm{i}\nu_\ell \to -\mathrm{i}\nu_\ell$, 
$\mathcal{M}^q_{11}$ is even only under $\mathbf{q} \to -\mathbf{q}$.
In the presence of a single band, these expressions recover the usual 
results~\cite{iskin20b, engelbrecht97}.

Next we introduce a unitary transformation
$
\Lambda_q = (\lambda_q + \mathrm{i} \theta_q)/\sqrt{2},
$
and associate $\lambda_q$ and $\theta_q$ with the amplitude and phase 
degrees of freedom, respectively. Assuming these are real functions in 
real space and time, we set $\lambda_{-q} = \lambda_q^*$ and 
$\theta_{-q} = \theta_q^*$, and express the effective action in the form
\begin{align}
\sum_q \left( \lambda_q^* \, \theta_q^* \right)
\left( \begin{array}{cc}
\mathcal{M}^{q,E}_{11} + \mathcal{M}^q_{12} & \mathrm{i} \mathcal{M}^{q,O}_{11} \\
- \mathrm{i} \mathcal{M}^{q,O}_{11} & \mathcal{M}^{q,E}_{11} - \mathcal{M}^q_{12}
\end{array} \right)
\left( \begin{array}{c}
\lambda_q \\ \theta_q
\end{array} \right),
\end{align}
where 
$
\mathcal{M}^{q,E}_{11} = (\mathcal{M}^q_{11} + \mathcal{M}^q_{22})/2
$
is an even function of $\mathrm{i}\nu_n$ and 
$
\mathcal{M}^{q,O}_{11} = (\mathcal{M}^q_{11} - \mathcal{M}^q_{22})/2
$
is an odd one. In particular when $q \to q_0 = (\mathbf{0}, 0)$, since the 
off-diagonal terms $\mathcal{M}^{q_0,O}_{11} = 0$ necessarily vanish, 
we observe that the amplitude and phase modes are always decoupled in 
the low-momentum and low-frequency limit. Furthermore the fact that 
$\mathcal{M}^{q_0,E}_{11} - \mathcal{M}^{q_0}_{12} = 0$ vanish, 
due to the saddle-point condition, 
(i.e., the order parameter Eq.~(\ref{eqn:op})), suggests that the 
low-frequency phase mode is always gapless, and we identify it as the 
Goldstone mode. Similarly, when $q \to q_I = (\mathbf{0}, 2\Delta_0)$, the fact that 
$\mathcal{M}^{q_I,E}_{11} + \mathcal{M}^{q_I}_{12} = 0$ vanish, 
due again to the saddle-point condition, 
suggests that the amplitude mode is gapped with $2\Delta_0$, and we 
identify it as the Higgs mode. Note that, since $\mathcal{M}^{q_I,O}_{11}$ 
does not vanish in general, the latter statement is strictly valid only 
in the BCS limit where $\mathcal{M}^{q_I,O}_{11}$ is small and negligible.

Given that the terms with the prefactor $(f-f')$ have the usual 
Landau singularity for $q \to (\mathbf{0}, 0)$ limit and causes the 
collective modes to decay, a small $q$ expansion is well-defined 
only in two cases: ($i$) just below $T_c$ as discussed in Ref.~\cite{iskin23}, 
and ($ii$) at $T = 0$ which is discussed in the main text.

\section{Numerical Implementation}
\label{sec:ni}

To verify the accuracy of our numerical results, we conducted calculations 
using two independent numerical implementations. In the first approach, 
we computed $\mathbf{k}$-space sums by dividing the BZ into approximately 
$10^6$ unit cells. In the second method, we converted $\mathbf{k}$-space 
sums into $\mathbf{k}$-space integrals through 
$
\sum_\mathbf{k} \to \frac{\mathcal{V}}{8\pi^3} \int d^3\mathbf{k}
$
and evaluated them using the adaptive CUBPACK integration library~\cite{cubpack}. 
Although both implementations yielded identical results, the second 
method speeds up the calculations significantly. 
Additionally, for any given values of $F$ and $U$, we obtained self-consistent 
solutions for Eqs.~(\ref{eqn:op}) and~(\ref{eqn:filling}) numerically 
by iterating $\Delta_0$ and $\mu$ through a hybrid root-finding 
algorithm of MINPACK that combines the bisection and secant methods. 
In particular, for generating the colored maps of various physical 
observables in the main text, we scanned the $(F, U)$ plane using 
a $400 \times 400$ mesh, resulting in approximately 160,000 repetitions 
of the iterative approach. Such a dense was necessary to reveal 
the DOS features on the periphery of the white regions including the flat 
bands at $\mu = -2t$, van Hove singularities at $\mu = \{0, 4t\}$ or 
equivalently at $F \approx \{1.19, 1.81\}$, and the critical semi-metal 
point at $\mu = 2t$ or equivalently at $F = 3/2$. 
While the iterative approach efficiently converged over a wide range 
of parameters, we encountered convergence issues only in the BCS 
regime when $\Delta_0/t \ll 1$. We emphasize that it is possible to address 
such convergence problems on demand, i.e., for any desired $F$ value, 
by providing more accurate input parameters for $\Delta_0$ and $\mu$ for 
the initial iteration. However, automating such fine-tuning operations 
proved to be difficult for the entire regime of interest in this paper.

\bibliography{refs}

\end{document}